\documentclass[epj]{svjour}
\usepackage{graphicx,color,amssymb}
\usepackage[numbers,sort&compress,merge]{natbib}

\newcommand{\lsim}{\raisebox{-0.13cm}{~\shortstack{$<$ \\[-0.07cm] $\sim$}}~} 
\newcommand{\gsim}{\raisebox{-0.13cm}{~\shortstack{$>$ \\[-0.07cm] $\sim$}}~}

\newcommand{\beq}{\begin{eqnarray}} 
\newcommand{\eeq}{\end{eqnarray}}

\begin{document}

\pagestyle{plain}
\title{Higgs Production in Neutralino Decays in the MSSM} 
\subtitle{The LHC and a Future $e^+e^-$ Collider}
\author{A. Arbey\inst{1}\inst{2}\inst{3} \and 
        M. Battaglia\inst{3} \inst{4} \and
        F. Mahmoudi\inst{1}\inst{2}\inst{3}\inst{5}% etc
}                     % Do not remove

\institute{
Universit{\' e} de Lyon, Universit{\' e} Lyon 1,
Centre de Recherche Astrophysique de Lyon, Saint-Genis Laval Cedex, F-69561, France; CNRS, UMR 5574 \and
Ecole Normale Sup{\' e}rieure de Lyon, France
\and
CERN, CH-1211 Geneva 23, Switzerland
\and
Santa Cruz Institute of Particle Physics, University of California, Santa Cruz,
CA 95064, USA \and
Clermont Universit\'e, Universit\'e Blaise Pascal, CNRS/IN2P3, LPC, BP 10448, 63000 Clermont-Ferrand, France}
\date{}

\abstract{
The search for the production of weakly-interacting SUSY particles at the LHC is crucial for testing supersymmetry 
in relation to dark matter. Decays of neutralinos into Higgs bosons occur over some significant part of the SUSY 
parameter space and represent the most important source of $h$ boson production in SUSY decay chains in the MSSM.
We study $h$ production in neutralino decays using scans of the phenomenological MSSM.
Whilst in constrained MSSM scenarios the decay $\tilde{\chi}^{0}_{2} \to h \tilde{\chi}^{0}_{1}$ is the dominant channel, this does not hold in more general MSSM scenarios. 
On the other hand, the  $\tilde{\chi}^0_{2,3} \to h \tilde{\chi}^{0}_{1}$ decays remain important 
and are highly complementary to multi-lepton final states in the LHC searches. The perspectives for the LHC analyses 
at 8 and 14~TeV as well as the reach of an $e^+e^-$ collider with $\sqrt{s}$ = 0.5, 1, 1.5 and 3~TeV are discussed.  
 % end of PACS codes
} %end of abstract
\maketitle
% main text

\section{Introduction}

The discovery of a Higgs boson by the ATLAS and CMS experiments~\cite{ATLAS:2012zz,CMS:2012zz} at the CERN LHC has opened 
an era of detailed studies of its production and decay properties. In particular, establishing if the discovered 
particle is the Standard Model (SM) Higgs boson or the manifestation of an extended Higgs sector is a key question. 
If the observed Higgs particle is the lightest Higgs boson, $h$, of a supersymmetric extension of 
the SM (SUSY), it could be significantly produced also in the decay chains of supersymmetric particles.
We shall show here that the decays of neutralinos, $\tilde{\chi}^0_j \to \tilde{\chi}^0_i h$ are the most important
process leading to the production of a $h$ boson in a SUSY decay chain in the MSSM. 
There have been several phenomenological studies of $h$ production in neutralino decays in various constrained
supersymmetric models~\cite{Gunion:1987kg,Gunion:1987yh,Bartl:1988cn,Djouadi:2001fa,Datta:2003iz,Huitu:2008sa,Gori:2011hj,Stal:2011cz,Howe:2012xe}, 
including some detailed assessments of the LHC potential~\cite{Kribs:2009yh,Kribs:2010hp,Datta:2003iz,Ghosh:2012mc}.
Results from searches for neutralino and chargino production conducted by ATLAS and CMS at the LHC have been reported
for the two~\cite{Atlas:2012gg,Chatrchyan:2012pka,Aad:2014vma,CMS-PAS-SUS-13-006} and 
three~\cite{Atlas:2012ku, Chatrchyan:2012pka,Aad:2014nua,CMS-PAS-SUS-13-006}, lepton channels as well as the
$bb \ell$ + MET channel~\cite{ATLAS-CONF-2013-093,CMS-PAS-SUS-13-017}, sensitive to $h$ production in neutralino decays.
This paper discusses the regions of the MSSM parameters where these decays into the lightest Higgs boson are relevant and 
the perspectives for their search at the LHC and a future lepton collider in the framework of the phenomenological minimal 
supersymmetric extension of the SM (pMSSM) with 19 free parameters~\cite{Djouadi:1998di}. This framework provides sufficient
freedom to the masses and couplings to explore the supersymmetric parameter space in a largely unbiased way and has been
adopted in phenomenological~\cite{AbdusSalam:2009qd,Sekmen:2011cz,Arbey:2011un,Arbey:2012dq,CahillRowley:2012kx,Arbey:2012bp}
and experimental~\cite{CMS:2013rda,CMS:2014mia} studies.
The study of the pMSSM parameter space with high statistics, flat scans of its parameters is briefly described in section~2.
Results from measurements and searches at LEP and LEP2, flavour physics and dark matter experiments provide already powerful
constraints. These are discussed in section~3 together with the implications of the LHC results from the search for SUSY particles
and the first determination of the properties of the discovered boson. Section~4 presents the regions of the viable pMSSM parameter
space probed by the different channels of neutralino decays, in particular those involving the lightest Higgs bosons.
In section~5, we contrast the results for the pMSSM with those obtained with the highly constrained, 5-parameters
CMSSM~\cite{Kane:1993td,Ellis:2002rp}, which have been widely studied in relation to dark
matter~\cite{Drees:1992am,Barger:1997kb,Baer:1997ai,Roszkowski:2001sb,Ellis:2012aa,Baer:2012uya}
and for benchmarking~\cite{Allanach:2002nj,Battaglia:2003ab,AbdusSalam:2011fc}.
Section 6 discusses the perspectives for the searches at LHC and a lepton collider, for various centre of mass energies
while section 7 has the conclusions.

\section{MSSM Scans and Tools}

This study is based on the analysis of a large sample of MSSM points obtained 
through a flat scan of the pMSSM parameters, which are varied in an uncorrelated way within the following ranges:
\begin{eqnarray}
1\leq \tan \beta \leq 60 \, , ~~~~~~~~~~~ \,  \nonumber \\
50~{\rm GeV} \leq M_A \leq 2~{\rm TeV}\, , ~~~~~~~~ \nonumber \\ 
\ -10~{\rm TeV} \leq A_f \leq 10 ~{\rm TeV} \, , ~~~~~~    \nonumber \\
~~~50~{\rm GeV} \leq M_{\tilde f_L}, M_{\tilde f_R}, M_3 
\leq 3.5~{\rm TeV}\, , \nonumber \\  
50~{\rm GeV} \leq M_1, M_2, |\mu| \leq 3.0~{\rm TeV}
\label{scan-range}
\end{eqnarray}
to generate a total of $2 \times 10^{8}$  pMSSM points. We also perform a scan of the CMSSM, in order to contrast
the results for the pMSSM with those of a highly constrained model.
For this we vary the CMSSM parameters within the following ranges:
\begin{center} 
\begin{eqnarray}
~~~~1\leq \tan \beta \leq 60 \, , ~~~~~ \,  \nonumber \\
~~~~~~\ -10~{\rm TeV} \leq A_0 \leq 10 ~{\rm TeV} \, , ~~  \nonumber \\
~~~0 \leq M_{0}, M_{1/2} \leq 3.5~{\rm TeV}. ~ 
\label{cscan-range}
\end{eqnarray}
\end{center}

In all scans, we select the generated points which fulfil a set of constraints derived from flavour physics 
and lower energy searches at LEP2 and the Tevatron, as summarised below and discussed in more details in 
Ref.~\cite{Arbey:2011un}, to which we refer also for further details on the pMSSM scans. These selected 
points are referred to as ``accepted points''  in the following. The pMSSM scans employ a number of programs 
and software tools. Only those most relevant to this study, are mentioned here, while further details are 
given in~\cite{Arbey:2011un}. SUSY mass spectra are generated with {\tt SOFTSUSY 3.2.3}~\cite{Allanach:2001kg}. 
Decay branching fractions are calculated using {\tt SDecay}~\cite{Muhlleitner:2003vg,Djouadi:2006bz}.
Higgs decay branching fractions are calculated with {\tt HDECAY (5.0)}~\cite{Djouadi:1997yw}. 
The flavour observables and dark matter relic density are calculated with {\tt SuperIso Relic v3.2}~\cite{flavor}. 
Cross sections are computed with {\tt Pythia 6.424}~\cite{Sjostrand:2006za}, also used for event generation. 
The NLO k-factors are evaluated using {\tt Prospino v2}~\cite{Beenakker:1996ed}. The physics object 
response for the LHC analyses is simulated using the {\tt Delphes 3.0} fast simulation 
package~\cite{Ovyn:2009tx,deFavereau:2013fsa}, tuned on the ATLAS detector performance and validated for the 
ATLAS multi-lepton neutralino and chargino analyses.

\section{Constraints}

We apply constraints from flavour physics, dark matter and SUSY searches at LEP, Tevatron and LHC. 
These have been discussed in details in Ref.~\cite{Arbey:2011un}.
In particular, we consider the decay $B_s \to \mu^+ \mu^-$, which can receive significant SUSY 
contributions at large values of $\tan\beta$. Here, we adopt the latest combined results from LHCb and CMS with the measurement of a branching fraction of BR($B_s^0 \to \mu^+ \mu^-) = (2.8 \pm 0.7)\times 10^{-9}$~\cite{Aaij:2013aka,Chatrchyan:2013bka,CMS:2014xfa}
from which we derive the constraints at 95\% C.L. after accounting for theoretical 
uncertainties. Dark matter also provides significant bounds for this analysis. We impose the constraint on the 
neutralino relic density of $10^{-4} < \Omega_{\chi} h^2 < 0.163$, derived from the PLANCK satellite 
result~\cite{Ade:2013zuv}, accounting for theoretical and cosmological uncertainties and 
allowing the neutralino be responsible for only part of the observed dark matter. 
In addition, we compare the sensitivity of the LHC searches to the new limits on dark matter direct detection for 
spin-independent $\tilde{\chi}$-nucleon scattering obtained from the XENON-100~\cite{Aprile:2012nq} and the 
LUX~\cite{Akerib:2013tjd} data.

Moreover, we apply the constraints in the Higgs sector by selecting only points having a mass of the lightest 
Higgs boson, $h$, in the mass range 123 $< M_h <$ 129~GeV and compatible with the constraints on the heavier MSSM 
Higgs bosons in the channel $H/A  \to \tau^+ \tau^-$~\cite{CMS-13-021}.

Searches for SUSY particles at the LHC in channels with missing $E_T$ have already provided a number of constraints 
relevant to this study, by excluding part of the  pMSSM phase space, corresponding to masses of the gluino and 
scalar quarks of the first two generations below $\sim 600$~GeV to 1~TeV and those of scalar bottom and top 
below $\sim 400$ to 800~GeV. Here, we use these results to compare the sensitivity of electroweak SUSY particle production 
to that of the ATLAS searches for scalar quarks of the first two generations and gluinos in the jets + MET 
channel~\cite{ATLAS-CONF-2013-047} and for scalar top and bottom quarks with $b$-tagged jets + MET~\cite{Aad:2013ija}.

\section{Neutralino decays to $h$ in the MSSM}

\begin{figure}[ht!]
\begin{center}
\begin{tabular}{cc}
\hspace*{-1mm}\includegraphics[width=0.265\textwidth]{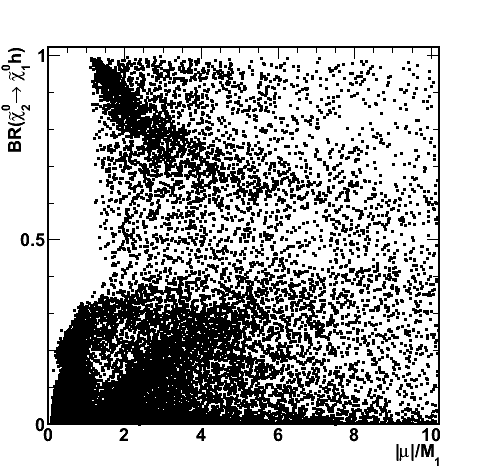} &
\hspace*{-8mm}\includegraphics[width=0.265\textwidth]{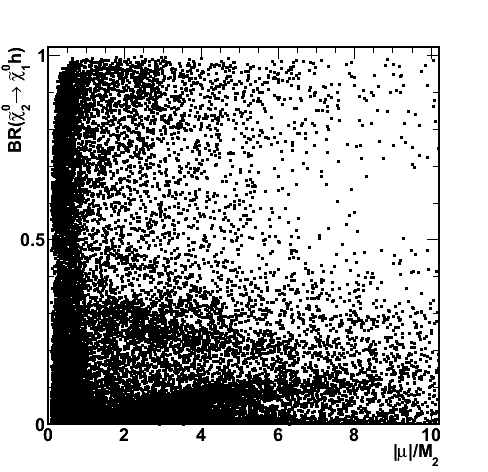} \\
\end{tabular}
\caption{The branching fraction $\tilde{\chi}^0_2 \to h \tilde{\chi}^0_1$ for accepted pMSSM points as a function of 
the ratio $|\mu|/M_1$ (left) and $|\mu|/M_2$ (right). The upper-left branches in both plots are due to bino-like $\tilde{\chi}^0_2$.} 
\label{fig:BRN2MuM12}
\end{center}
\end{figure}

Neutralino decays are the most important source of $h$ bosons in SUSY decays in the MSSM. The study of $h$ bosons appearing
in the decay of SUSY particles in models obtained by scanning the pMSSM parameter space shows that in 53\% of the cases
they originate from neutralino decays, either from direct production or from a scalar quark decay chain. This value slightly
exceeds that obtained for the occurence of chargino decays $\tilde{\chi}^{\pm}_{2} \rightarrow h \tilde{\chi}^{\pm}_{1}$ (45\%)
while the $\tilde{\tau}^{\pm}_{2} \rightarrow h \tilde{\tau}^{\pm}_{1}$,
$\tilde{t}_{2} \rightarrow h \tilde{t}_{1}$ and $\tilde{b}_{2} \rightarrow h \tilde{b}_{1}$ processes have only a marginal
contribution (2\%).
The occurence of $h$ bosons in SUSY decays characterises $\sim$1.5\% of the accepted pMSSM points considered in this study.  
In the MSSM the couplings of the Higgs bosons to charginos and neutralinos are maximal for higgsino--gaugino 
mixed states, while the gauge boson couplings to neutralinos are maximal for higgsino--like states. The 
$h$ couplings to neutralino are suppressed by powers of $|\mu|/M_2$ or $M_2/|\mu|$ in the gaugino--like and 
higgsino--like regions, respectively. The branching fraction of $\tilde{\chi}^0_2 \to h \tilde{\chi}^0_1$ for a set of
accepted pMSSM points is shown as a function of $|\mu|/M_1$ and $|\mu|/M_2$ in Figure~\ref{fig:BRN2MuM12}.
The larger values of this branching fraction are obtained mostly in the region where $|\mu| \lesssim M_2$ and the neutralino
is bino. On the contrary, in the region where $|\mu| \lesssim M_1$ the branching ratio is limited to no more than 30\%.

In the case the charginos and neutralinos are gaugino--like (i.e.\ when the higgsino mass parameter is much larger 
than the wino mass parameter, $|\mu| \gg M_2$) or higgsino--like (i.e. in the opposite situation $|\mu| \ll M_2$), 
this results into the dominance of the decays of the heavier charginos and neutralinos into the lighter states and 
Higgs bosons, over the same decays with gauge boson final states~\cite{Djouadi:2001fa,Datta:2003iz}. 
This is also the case for the so-called ``little cascade" in the region $M_2 \gsim M_1 \gg |\mu|$ where the 
branching fraction for the decay of the $\tilde{\chi}_2^0$ into the LSP neutralino and the lighter $h$ boson 
$\tilde{\chi}_2^0 \to h \tilde{\chi}_1^0$, is in general larger than that for the decay 
$\tilde{\chi}_2^0 \to Z \tilde{\chi}_1^0$, when kinematically accessible in the two-body 
channel \footnote{In the three--body process, the $h$ has to be virtual and the rate is suppressed by the small couplings 
of the $h$ state to light fermions.}. Thus, the decay $\tilde{\chi}_2^0 \to h \tilde{\chi}_1^0$ is important in these 
regions, unless sleptons are light and the additional channels where $\tilde{\chi}_2^0$ decays into a lepton and its 
super-partner are open. These decays are most important if the neutralino is gaugino--like, since the coupling to the 
higgsino component is suppressed due to the small lepton mass.

\begin{figure}[ht!]
\begin{center}
\begin{tabular}{cc}
\hspace*{-2mm}\includegraphics[width=0.265\textwidth]{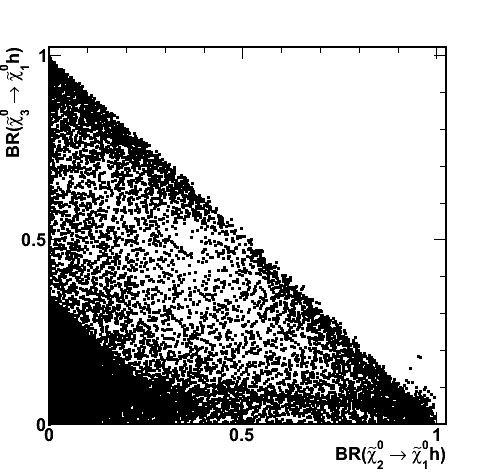} &
\hspace*{-7mm}\includegraphics[width=0.265\textwidth]{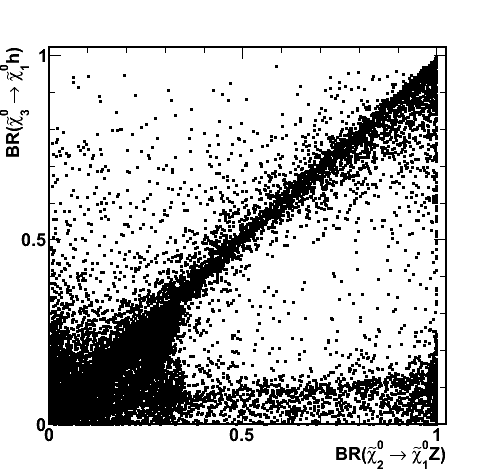} \\
\end{tabular}
\caption{The branching fraction $\tilde{\chi}^0_3 \to h \tilde{\chi}^0_1$ as a function of that for  
$\tilde{\chi}^0_2 \to h \tilde{\chi}^0_1$ (left) and $\tilde{\chi}^0_3 \to h \tilde{\chi}^0_1$ as a function of that for 
$\tilde{\chi}^0_2 \to Z \tilde{\chi}^0_1$ (right) for accepted pMSSM points, showing the complementarity of the decays 
of $\tilde{\chi}^0_3$ and $\tilde{\chi}^0_2$ to $h$ and $Z$.}
\label{fig:BRN23}
\end{center}
\end{figure}

It is important to observe that for small values of $||\mu| - M_2|$, where the contribution of the decay into $h$ is 
enhanced, the mass difference $M_{\chi^0_3} - M_{\chi^0_2}$ is small, typically of the order of 50~GeV or less. 
In this regime, the production cross sections for $\tilde{\chi}^0_2 \tilde{\chi}^{\pm}_1$ and 
$\tilde{\chi}^0_3 \tilde{\chi}^{\pm}_1$ in $pp$ collisions 
are comparable. Due to the nature of the $\tilde{\chi}^0_2$ and $\tilde{\chi}^0_3$ states, the branching fractions into $h$ 
and $Z$ of these two neutralinos are complementary, i.e.\ the yield into $h$ for the $\tilde{\chi}^0_2$ is approximately 
equal to that into $Z$ of the $\tilde{\chi}^0_3$ states, and vice versa. 
This is illustrated in Figure~\ref{fig:BRN23} which shows the branching fraction of $\tilde{\chi}^0_3 \to h \tilde{\chi}^0_1$ as a function of those for $\tilde{\chi}^0_2 \to h \tilde{\chi}^0_1$ and $\tilde{\chi}^0_2 \to Z \tilde{\chi}^0_1$. The left panel reveals the anti-correlation between the $\tilde{\chi}^0_2$ and $\tilde{\chi}^0_3$ decays, while that on the right shows the linear correlation between $\tilde{\chi}^0_3 \to h \tilde{\chi}^0_1$ and $\tilde{\chi}^0_2 \to Z \tilde{\chi}^0_1$ decays for most of the accepted points. This highlights the complementarity between the search for $\tilde{\chi}^0_2$ and
$\tilde{\chi}^0_3$ with decays into $h$ and $Z$ and the interest in pursuing both in the LHC analyses.

\begin{figure}[t!]
\begin{center}
\includegraphics[width=0.30\textwidth]{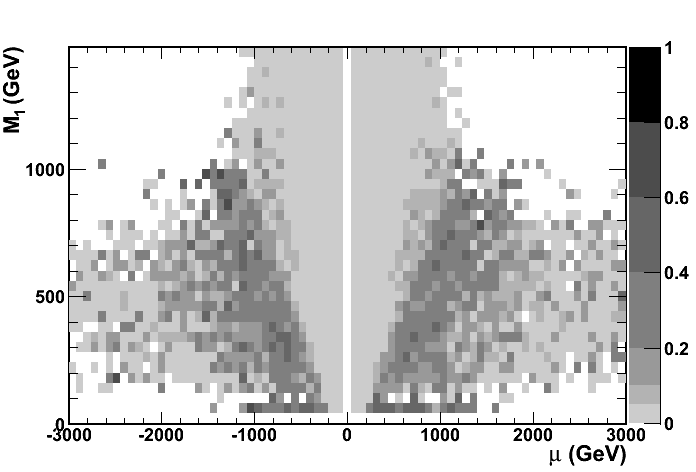}\\
\includegraphics[width=0.30\textwidth]{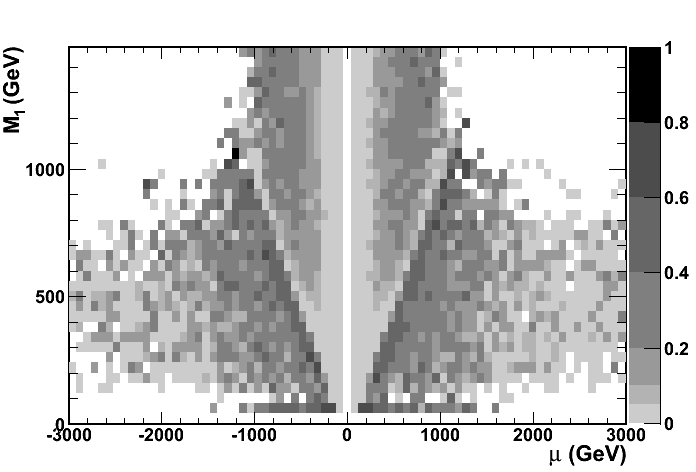} \\
\includegraphics[width=0.30\textwidth]{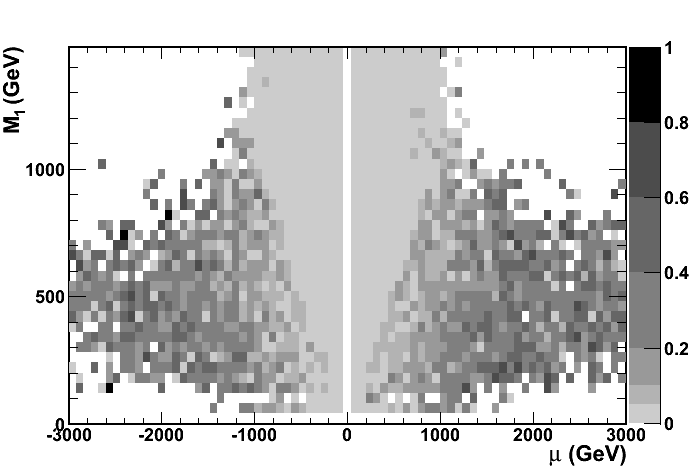}\\
\includegraphics[width=0.30\textwidth]{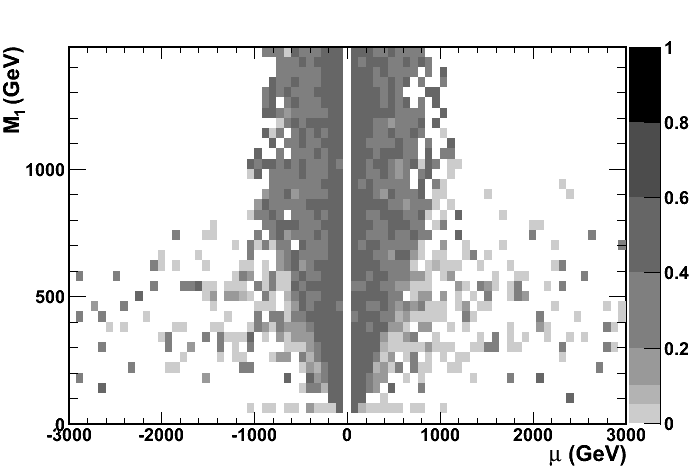} 
\caption{Fraction of accepted pMSSM points in the $[\mu - M_1]$ MSSM plane having 
$\tilde{\chi}^0_2 \to h \tilde{\chi}^0_1$ (top) $Z$ (upper), $\tilde{\ell} \ell$ or 
$\tilde{\tau} \tau$ (center) and three-body (lower) as dominant decay.}
\label{fig:M1MuN2}
\end{center}
\end{figure}

\begin{figure}[ht!]
\begin{center}
\includegraphics[width=0.30\textwidth]{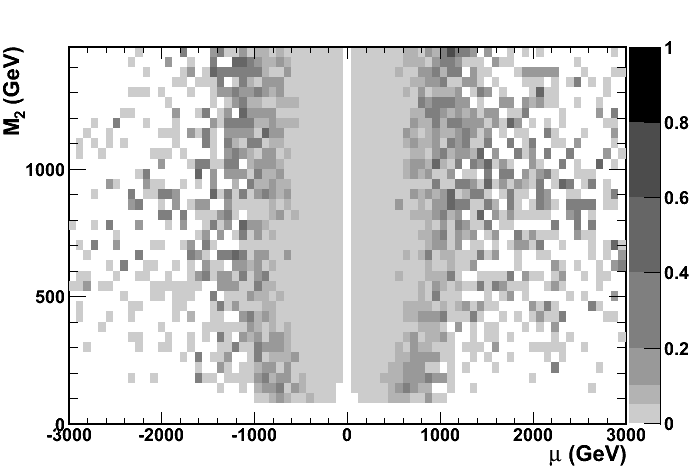} \\
\includegraphics[width=0.30\textwidth]{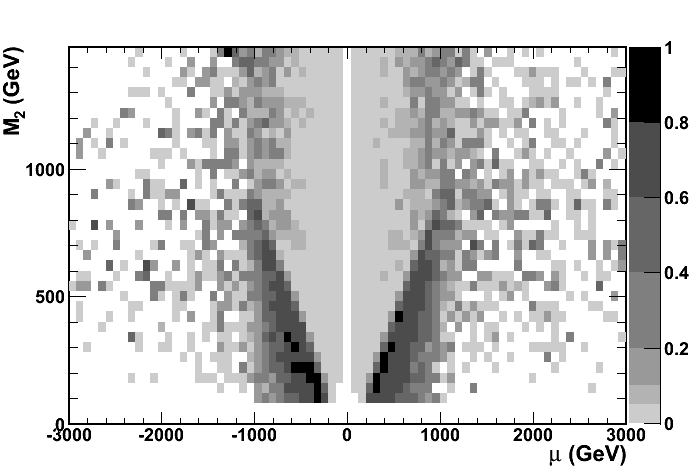} \\
\includegraphics[width=0.30\textwidth]{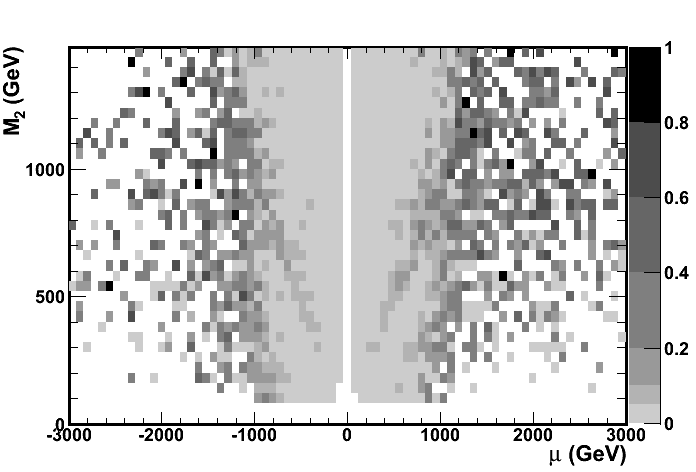} \\
\includegraphics[width=0.30\textwidth]{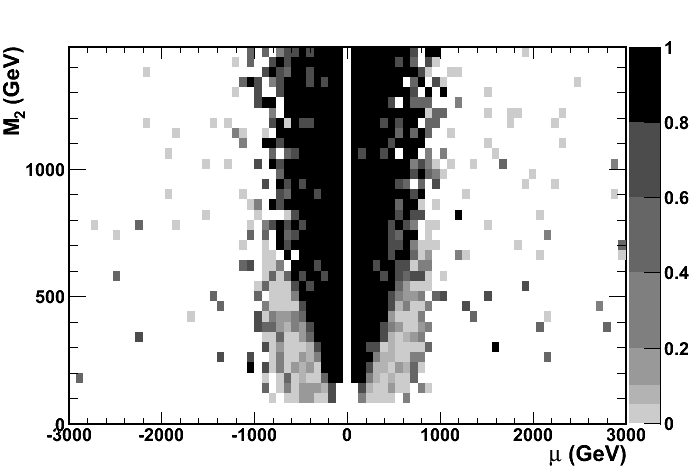} 
\caption{Same as Figure~\ref{fig:M1MuN2} for the $[\mu - M_2]$ MSSM plane with the fraction of accepted pMSSM points having $\tilde{\chi}^0_2 \to h \tilde{\chi}^0_1$ (top) $Z$ (upper), $\tilde{\ell} \ell$ or $\tilde{\tau} \tau$ (lower) and three-body (bottom) as dominant decay.}
\label{fig:M2MuN2}
\end{center}
\end{figure}

We study the regions explorable by the $\tilde{\chi}^0_{2,3} \to h \tilde{\chi}^0_1$ decay using our set of accepted 
pMSSM points. The extension of these regions depend only on a subset of the pMSSM parameters, mostly
$M_1$, $M_2$, $\mu$, $M_2-M(\tilde{e}_{L,R})$. However, the constraints discussed in the previous section introduce
non-trivial correlations with the other pMSSM parameters and make advantageous to perform this study by keeping all the
parameters free. In order to characterise the occurrence of the various decays in different regions of the parameter space,
we first study the fraction of the accepted pMSSM points into each bin in the $[\mu, M_1]$ and $[\mu, M_2]$ planes where 
$h \tilde{\chi}^0_1$, $Z \tilde{\chi}^0_1$ and the sum of $\tilde{\ell} \ell$ and $\tilde{\tau} \tau$ is the dominant 
two-body neutralino decay mode. We define the dominant decay mode as that having a branching fraction larger than both 
0.2 and any of the other two-body decay channels. Results are shown for $\tilde{\chi}^0_2$ in Figures~\ref{fig:M1MuN2} 
and \ref{fig:M2MuN2}, where we also show the region where the sum of branching fractions for the three-body decays exceeds 
both 0.2 and any of the individual two-body processes. While the exact values for the fraction of points depend on the 
parameter ranges adopted for our scans, given in Eq.(\ref{scan-range}), the location of the parameter regions yielding
decays into $h$ bosons from Figure~\ref{fig:M2MuN2} is of general validity. 

As shown in Figure~\ref{fig:M1MuN2}, the $h \tilde{\chi}_2^0 \tilde{\chi}_1^0$ couplings are larger than the 
$Z \tilde{\chi}_2^0 \tilde{\chi}_1^0$ whenever $M_1 \lsim |\mu|$ and the $\tilde{\chi}_1^0$ is mostly bino.
The region 
where $|\mu|  \approx M_1$ is excluded as the decay $\tilde{\chi}_2^0 \to h \tilde{\chi}_1^0$ is not kinematically allowed.
Very small values of $|\mu|$ are excluded from the searches for charginos at LEP2, which implies 
that $|\mu| \gsim$ 90~GeV. The $\tilde{\chi}_2^0 \to Z \tilde{\chi}_1^0$ mode is dominant for low values of $\mu$ where 
the LSP is higgsino--like and, thus, there is  a strong coupling of the $Z$ boson to $\tilde{\chi}_2^0 \tilde{\chi}_1^0$ 
states. There is a line $M_Z \lsim M_{\chi_2^0} - M_{\chi_1^0} \lsim M_h$ in which the decay channel 
$\tilde{\chi}_2^0 \to h \tilde{\chi}_1^0$ is kinematically closed and the decay $\tilde{\chi}_2^0 \to Z \tilde{\chi}_1^0$ 
becomes dominant, provided that the sleptons are not light enough for the $\tilde{\ell} \ell$ channel to contribute.
Finally, at small $|\mu|$ values, $\tilde{\chi}^0_1$ and $\tilde{\chi}^0_2$ are higgsino--like, but the splitting is 
typically small, so the two-body decays through $h$ or $Z$ are forbidden and the three-body modes dominate.
It is important to point out that the parameters which increase $\tilde{\chi}_{2,3}^0 \to h \tilde{\chi}_1^0$ 
do not cause an increase of the $h$ decays into $\tilde{\chi}^0_1 \tilde{\chi}^0_1$ pairs and the branching fractions
for the two processes are largely uncorrelated. Therefore, the rejection of the pMSSM points giving invisible Higgs decay
rates above 15\% does not reduce the occurence of neutralino decays into the lightest Higgs boson.   
The dependence of the dominant channel on $M_2$ is less pronounced, as shown in  Figure~\ref{fig:M2MuN2}. 
In the region where $M_2 \lsim |\mu|$, the $\tilde{\chi}^0_2 \to Z \tilde{\chi}^0_1$ is dominant, while the three-body
channels become very important for $|\mu| \lsim M_2$ due to the small mass splitting between the two light higgsinos.

Through the decays $h \to b \bar b$, $h \to \tau^+ \tau^-$, $Z \to b \bar b$, $Z \to \tau^+ \tau^-$, these 
processes lead to the final states $b \bar b \ell$ + $E_{T}^{\mathrm{missing}}$ and $\tau \tau \ell$ + 
$E_{T}^{\mathrm{missing}}$ at the LHC, where the $\tilde{\chi}^0_{2,3}$ is produced in association with a chargino 
decaying $\tilde{\chi}^{\pm}_1 \to W^{\pm} \tilde{\chi}^0_1 \to \ell^{\pm} \nu \tilde{\chi}^0_1$.
An alternative, intriguing scenario leading to the same $\tau \tau \ell$ + $E_{T}^{\mathrm{missing}}$ final state 
arises if the $\tilde \tau_1$ is light. In this case, the decay 
$\tilde{\chi}^0_2 \to \tilde \tau_1 \tau \to \tau \tau \tilde{\chi}^0_1$ may be enhanced. 

\section{Neutralino decays to $h$: pMSSM to CMSSM comparison}

\begin{figure}[h!]
\begin{center}
\begin{tabular}{cc}
\hspace*{-3mm}\includegraphics[width=0.270\textwidth]{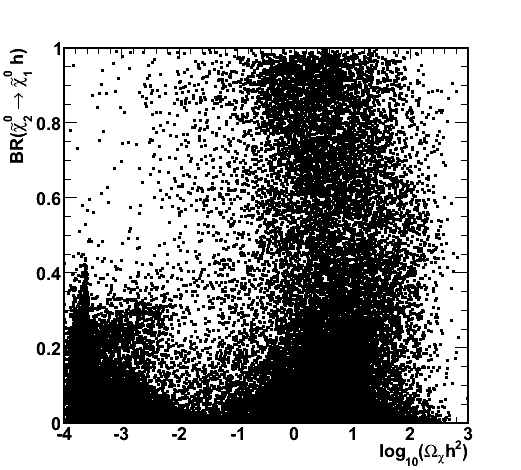} &
\hspace*{-8mm}\includegraphics[width=0.263\textwidth]{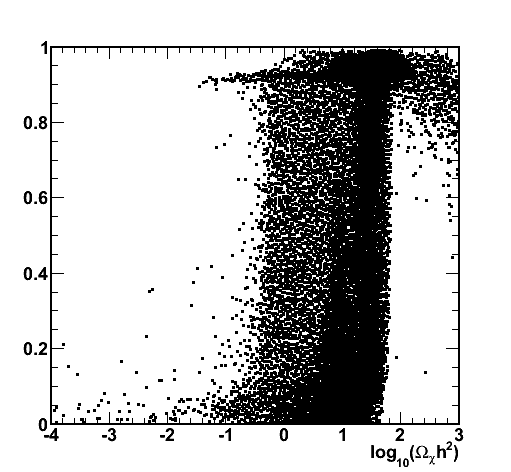} \\
\end{tabular}
\caption{Branching fraction for $\tilde{\chi}^0_2 \to  h \tilde{\chi}^0_1$ as a function of the neutralino relic density, 
$\Omega_{\chi} h^2$, for pMSSM (left) and CMSSM (right) points.}
\label{fig:BRN2HOh2}
\end{center}
\end{figure}

$W h$ production from $\tilde{\chi}^{\pm}_1 \tilde{\chi}^0_2$ decays at the LHC has been discussed in the context 
of the CMSSM in Ref.~\cite{Baer:2012ts}, where it is found that the decay  $\tilde{\chi}^0_2 \to h \tilde{\chi}^0_1$ 
is dominant for typical choices of the model parameters. In our analysis we find indeed that 
$\tilde{\chi}^0_2 \to h \tilde{\chi}^0_1$ 
is the dominant decay, with a branching fraction in excess of 0.20, for 71\% of the accepted CMSSM points before
applying the relic density constraint. However, this is no longer the case when we impose the full set of constraints.
Of the accepted points with 
$\tilde{\chi}^0_2 \to h \tilde{\chi}^0_1$ as the dominant decay, only 10$^{-4}$ fulfil the constraint 
$10^{-4} < \Omega_{\chi} h^2 < 0.163$ and $8 \times 10^{-5}$ the tighter requirement of $0.076 < \Omega_{\chi} h^2 < 0.163$ 
set by the PLANCK cosmic microwave background (CMB) bound.
In fact, the region where $\tilde{\chi}^0_2 \to h \tilde{\chi}^0_1$ is dominant corresponds to values of neutralino 
relic density, $\Omega_{\chi} h^2$, which largely exceed the upper limit of the PLANCK result~\cite{Ade:2013zuv}, 
as shown in Figure~\ref{fig:BRN2HOh2}.
This is due to the peculiar nature of the CMSSM, which has large $\mu$ values, with $|\mu| \gg M_2$ making the neutralinos 
to be gaugino--like, as observed above. The general prevalence of the $\tilde{\chi}^0_2 \to h \tilde{\chi}^0_1$ decay, 
observed in the CMSSM before the $\Omega_{\chi} h^2$ constraint, cannot be generalised to the MSSM. But in turn the greater 
flexibility of the pMSSM allows to reconcile the relic density constraint to large coupling of the $h$ to the neutralinos 
(see Figure~\ref{fig:BRN2HOh2}). 
For comparison, in the pMSSM the decay into $\tilde{\chi}^0_1 h$ is dominant in only 10\% of the accepted points before 
the relic density constraint. But now 28\% of these fulfil also the loose relic density constraint of 
$10^{-4} < \Omega_{\chi} h^2 < 0.163$ and still 2.2\% make also the tighter requirement of 
$0.076 < \Omega_{\chi} h^2 < 0.163$. These points are almost all in the so-called ``$A^0$ annihilation funnel'', where
neutralinos in the early universe annihilate through the $A^0$ pole to acquire a relic density in agreement with the CMB
data~\cite{Griest:1990kh,Ellis:2001msa}.
It is therefore important to reconsider the apparent dominance of the 
$\tilde{\chi}^0_2 \to h \tilde{\chi}^0_1$ decay in the CMSSM, when examining the SUSY phenomenology in the context of 
more general models, or even in the CMSSM itself after imposing the dark matter relic density constraint. 
Still, the process of $h$ production in neutralino decays is of crucial importance in specific regions of the parameter 
space, which are largely complementary to those yielding decays into sleptons or $Z$ bosons, probed by the multi-lepton 
analyses at the LHC.

\section{Neutralino decays to $h$: Experimental Searches}

In the following we discuss the perspectives for investigating these decays at the LHC and at an $e^+e^-$ linear 
collider (LC). We consider for the LHC an integrated luminosity of 25~fb$^{-1}$ at 8~TeV, 300~fb$^{-1}$ and 
3000~fb$^{-1}$ at 14~TeV. 
For the LC we consider 500~fb$^{-1}$ of integrated luminosity at $\sqrt{s}$ = 0.5~TeV and 1~ab$^{-1}$ at 1, 1.5 and 3~TeV. 
For each of these scenarios, we combine the decay branching fractions with the relevant neutralino production cross sections 
to study the number of expected events and compare them to the expected backgrounds. For the LHC, we compare the regions of 
exclusion from the $bb \ell$ mode to the two- and three-lepton channels. 
 
\subsection{LHC}

\begin{figure}[t!]
\begin{center}
\includegraphics[width=0.30\textwidth]{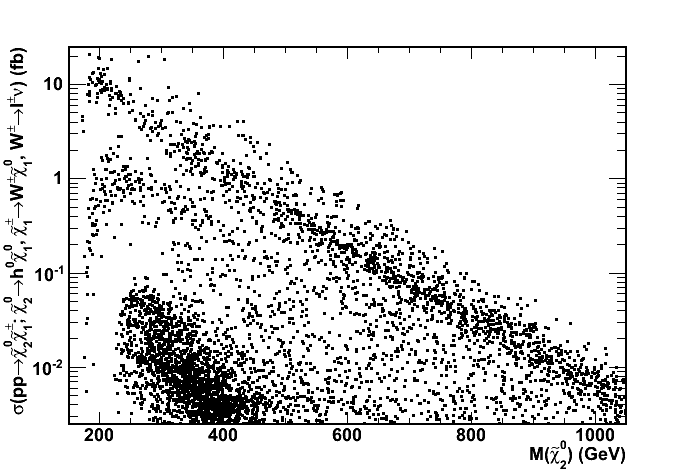} \\
\includegraphics[width=0.30\textwidth]{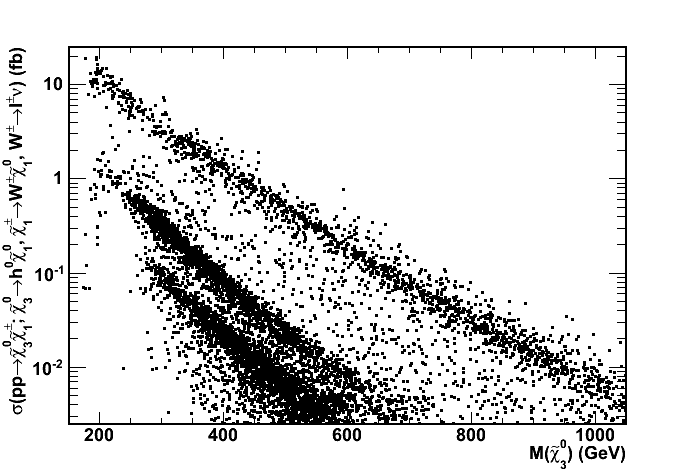}
\caption{Cross section for $\tilde{\chi}^{\pm}_1 \tilde{\chi}^0_2$ (upper) and $\tilde{\chi}^{\pm}_1 \tilde{\chi}^0_3$ 
(lower) production vs.\ $M_{\chi^0_2}$ and  $M_{\chi^0_3}$, respectively, with $\tilde{\chi}^{0}_{2, 3} \to h \tilde{\chi}^0_1$;
$h \to bb$, $\tau \tau$ and $\tilde{\chi}^{\pm}_1 \to W^{\pm} \tilde{\chi}^0_1$; $W^{\pm} \to \ell^{\pm} \nu$ in 8~TeV $pp$ 
collisions for a set of accepted pMSSM points.}
\label{fig:xsec8}
\end{center}
\end{figure}

\begin{figure}[h!]
\begin{center}
\includegraphics[width=0.30\textwidth]{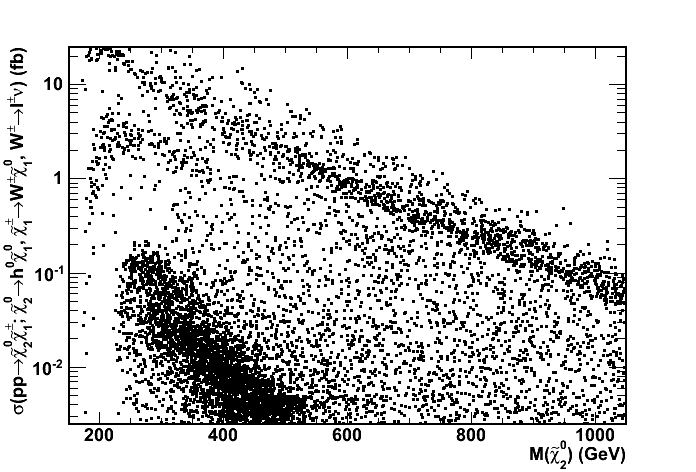} \\
\includegraphics[width=0.30\textwidth]{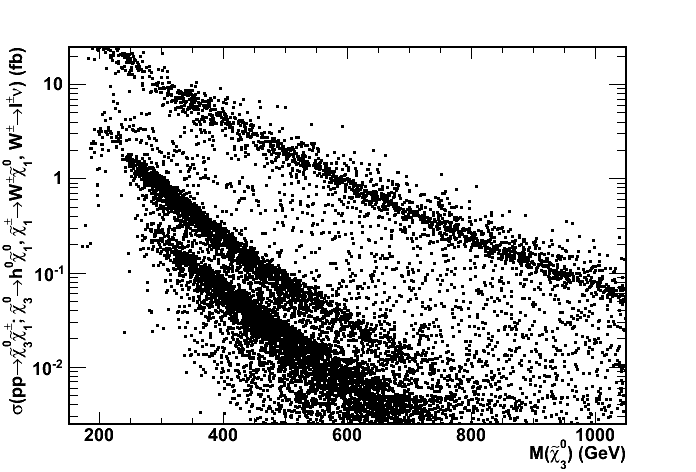}
\caption{Same as Figure~\ref{fig:xsec8} for 14~TeV $pp$ collisions.}
\label{fig:xsec14}
\end{center}
\end{figure}

As a first step, we combine the results on the branching fractions discussed in the previous section to the expected relevant 
cross sections for the associated chargino neutralino production. Through the decay 
$\tilde{\chi}^{\pm}_{1} \to W^{\pm} \tilde{\chi}^0_1$, $W^{\pm} \to \ell^{\pm} \nu$, this process produces a high $p_T$ lepton, 
which is very valuable for the event trigger and subsequent selection. The cross sections for chargino neutralino production at 
8 and 14~TeV, followed by $\tilde{\chi}^{0}_{2, 3} \to h \tilde{\chi}^0_1$; $h \to b \bar b$, $\tau \tau$ and 
$\tilde{\chi}^{\pm}_{1} \to W^{\pm} \tilde{\chi}^0_1$, $W^{\pm} \to \ell^{\pm} \nu$ are given in Figures~\ref{fig:xsec8} 
and ~\ref{fig:xsec14}, respectively. Because of the small values of these cross sections and the large spread for a given 
neutralino mass, luminosity is as important as energy for probing electroweak SUSY particle production at the LHC. The 
high-luminosity LHC program (HL-LHC) bringing a ten-fold increase in the data statistics will be extremely beneficial to 
these studies.

The ATLAS and CMS experiments have conducted searches for electroweak production of SUSY particles in the two- and three-lepton + MET 
channels and the $bb \ell$ + MET channel. Accepted pMSSM scan points fulfilling the selection criteria discussed above and having an 
inclusive chargino and neutralino production cross section yielding at least ten signal events at 8 or 14 TeV collision energies are 
studied in details. A set of 20k SUSY events with inclusive neutralino and chargino production is generated for each of these points at 
both centre-of-mass energies. Simulated events are passed through a fast parametric simulation using the {\tt Delphes 3.0} package which 
simulates the physics objects used in the subsequent analysis. Jets are reconstructed using the anti-kt algorithm~\cite{Cacciari:2008gp}, 
implemented in the {\tt FastJet} package~\cite{fastjet}. The selection criteria of the two- and three-lepton + MET~\cite{Aad:2014vma,Aad:2014nua} 
and the $bb \ell$ + MET~\cite{ATLAS-CONF-2013-093} analyses of ATLAS are applied. The exclusion of the pMSSM points is 
assessed using the CLs technique~\cite{Read:2002hq}, where the number of expected signal events is obtained from the result of the selection 
on the generated events and that of background events is taken from that estimated for the ATLAS analyses, appropriately rescaled to the 
assumed luminosity and collision energy. The results of this simulation are validated by comparing the 95~\% C.L.\ exclusion contours 
with those expected for the ATLAS analyses, under the same assumptions. These contours agree within 20-30\%. 

\begin{figure}[h!]
\begin{center}
\includegraphics[width=0.25\textwidth]{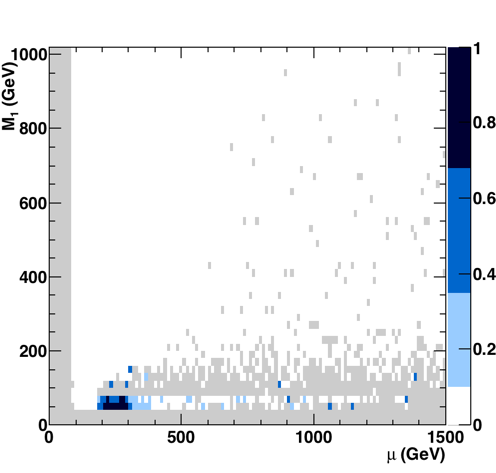} 
\includegraphics[width=0.25\textwidth]{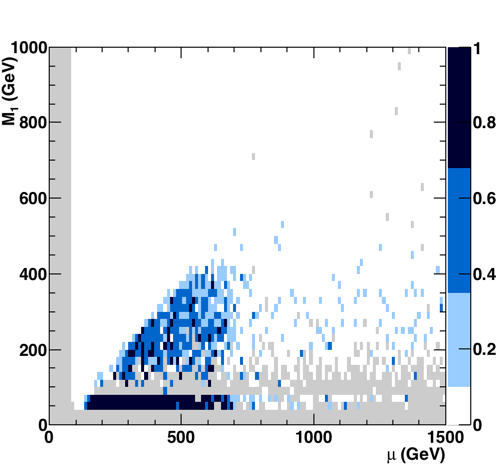} 
\includegraphics[width=0.25\textwidth]{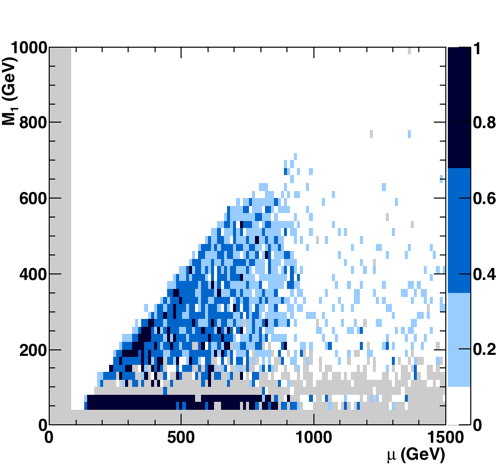}
\caption{Fraction of accepted pMSSM points in the $[|\mu| - |M_1|]$ plane excluded by the $bb \ell$ + MET channel at 8~TeV (upper) and their extrapolations at 14~TeV for 300~fb$^{-1}$ (centre) and  3000~fb$^{-1}$ (lower).}
\label{fig:HM1Mu}
\end{center}
\end{figure}

We study the fraction of accepted pMSSM points for which the different channels and their combination lead to a 95\% C.L. 
exclusion for 25~fb$^{-1}$ at 8~TeV, 300~fb$^{-1}$ and 3000~fb$^{-1}$ at 14~TeV. 
The fractions of the accepted pMSSM points excluded at 8 and 14~TeV by the 
$bb \ell$ + MET channel in the $[|\mu| - |M_1|]$ plane are given in Figure~\ref{fig:HM1Mu}. 
These fractions for the different channels over the full scan range are compared in Table~\ref{tab:summary}. Because the product of
the $\tilde{\chi}^{\pm}_1 \tilde{\chi}^0_{2,3}$ production cross section and decay branching fraction into $h$ is of $\cal{O}$(0.1~fb), or less, at
8~TeV, the Run-1 LHC searches have just scraped the region of interest for these processes in the MSSM. The Run-2 searches will definitely attain
an interesting sensitivity for neutralino masses below 600-700~GeV, as can be seen in Figure~\ref{fig:HM1Mu}.
The values of the fractions of pMSSM points depend on the ranges for the pMSSM parameters adopted in the scan, given at Eq.~(\ref{scan-range}). 
However, the location of the regions of larger sensitivity, highlighted by the large fractions of excluded pMSSM points, 
are of general validity. 
The MSSM parameter coverage provided by SUSY weak production searches at 8~TeV appears, in general, to be marginal. Of our accepted scan points 
with SUSY masses up to 3~TeV, only 0.3\% are excluded by the combinations of the three channels considered here. These are also independently 
excluded by searches in jets + MET and monojet channels. The factor of $\sim$3 increase of the 
$pp \to \tilde{\chi}^{\pm}_{1} \tilde{\chi}^{0}_{2, 3}$ production cross section and the factor of $\sim \!10-100$ increase in the 
assumed statistics, make the coverage of the parameter space to expand significantly when moving from 
25~fb$^{-1}$ at 8~TeV to 300~fb$^{-1}$ and 3000~fb$^{-1}$ at 14~TeV, with 3\% of the pMSSM points expected to be excluded with the HL-LHC data.

The leptonic channels are sensitive to the contribution of light sleptons in the decays 
$\tilde{\chi}^0_2 \to \tilde{\ell} \ell$; $\tilde \ell \to \ell \tilde{\chi}^0_1$, and $Z$ bosons in the decay 
$\tilde{\chi}^0_2 \to Z \tilde{\chi}^0_1$; $Z \to \ell \ell$. In the unconstrained MSSM, slepton masses can be pushed well 
above the $\tilde{\chi}^0_2$ mass, so that the only remaining dominant two-body $\tilde{\chi}^0_2$ decays are either 
$h \tilde{\chi}^0_1$ or $Z \tilde{\chi}^0_1$. This motivates the pursue of the study of the challenging $h \tilde{\chi}^0_1$ 
channel, through the $b \bar b \ell$ + $E_{T}^{\mathrm{missing}}$  topology, which complement the $Z \tilde{\chi}^0_1$ channel and
provides an increasing fraction of the LHC sensitivity to chargino-neutralino pair production as the energy and integrated luminosity
increase.  The neutralino decay into $h$ accounts for approximately one third of the excluded points with 25~fb$^{-1}$ at 8~TeV to more than
half of them for 3000~fb$^{-1}$ at 14~TeV.  The $h$ decay into $\tau \tau$ pairs also leads to the same
$\tau^+ \tau^- \ell$ + $E_{T}^{\mathrm{missing}}$ final state as $Z \tilde{\chi}^0_1$; $Z \to \tau \tau$. 
If a signal is observed in the $b \bar b \ell$ + $E_{T}^{\mathrm{missing}}$ 
and/or $\tau^+ \tau^- \ell$ + $E_{T}^{\mathrm{missing}}$ modes it would become essential to attempt to identify the contributing 
channels, possibly through a fit to the $bb$ invariant mass, as discussed below, and the $\tau \tau$ transverse mass.

% \begin{figure}[h!]
% \begin{center}
% \begin{tabular}{cc}
% \includegraphics[width=0.23\textwidth]{hM1M2-08TeV-25fb-Ewk.png} &
% \includegraphics[width=0.23\textwidth]{hM1M2-08TeV-25fb-bbl.png} \\
% \includegraphics[width=0.23\textwidth]{hM1M2-14TeV-300fb-Ewk.png} &
% \includegraphics[width=0.23\textwidth]{hM1M2-14TeV-300fb-bbl.png} \\
% \includegraphics[width=0.23\textwidth]{hM1M2-14TeV-3000fb-Ewk.png} &
% \includegraphics[width=0.23\textwidth]{hM1M2-14TeV-3000fb-bbl.png} \\
% \end{tabular}
% \caption{Fraction of accepted pMSSM points as in Figure~\ref{fig:HM1Mu} in the $[|M_1| - |M_2|]$ plane.}
% \label{fig:HM1M2}
% \end{center}
% \end{figure}

\begin{table}
\begin{center}
\begin{tabular}{|l|c|c|c|}
\hline
   & 8~TeV & 14 TeV & 14 TeV \\
   & 25~fb$^{-1}$ & 300~fb$^{-1}$ & 3000~fb$^{-1}$ \\
\hline
2/3 $\ell$ & 0.002 & 0.008 & 0.016 \\
3$\ell$(Z) & 0.001 & 0.004 & 0.009 \\
$bb$(h) $\ell$ & 0.001 & 0.009 & 0.015 \\
All EWK & 0.003 & 0.015 & 0.029\\
\hline
\end{tabular}
\caption{Fraction of pMSSM points excluded in the various channels at 8 and 14~TeV.}
\label{tab:summary}
\end{center}
\end{table}
The nature of the lightest neutralino has been analysed for the points excluded at 8 and 14~TeV and the results are summarised in 
Table~\ref{tab:n1}. As expected, the largest fraction of points has bino $\tilde{\chi}^0_1$, since this is dominant for 
light $\tilde{\chi}^0_1$ states, followed by higgsino--gaugino mixed states which become increasingly important as the search 
sensitivity expands with higher energy and larger data sets.
\begin{table}
\begin{center}
\begin{tabular}{|l|c|c|c|}
\hline
   & 8~TeV & 14 TeV & 14 TeV \\
   & 25~fb$^{-1}$ & 300~fb$^{-1}$ & 3000~fb$^{-1}$ \\
\hline
Bino LSP     & 0.789 & 0.706 & 0.679 \\
Wino LSP     & 0.0   & 0.0   & 0.010 \\
Higgsino LSP & 0.0   & 0.0   & 0.0   \\
Mixed LSP    & 0.211 & 0.294 & 0.311 \\
\hline
\end{tabular}
\caption{Neutralino LSP nature for the pMSSM points excluded at 8 and 14~TeV.}
\label{tab:n1}
\end{center}
\end{table}

\begin{figure}[h!]
\begin{center}
\includegraphics[width=0.35\textwidth]{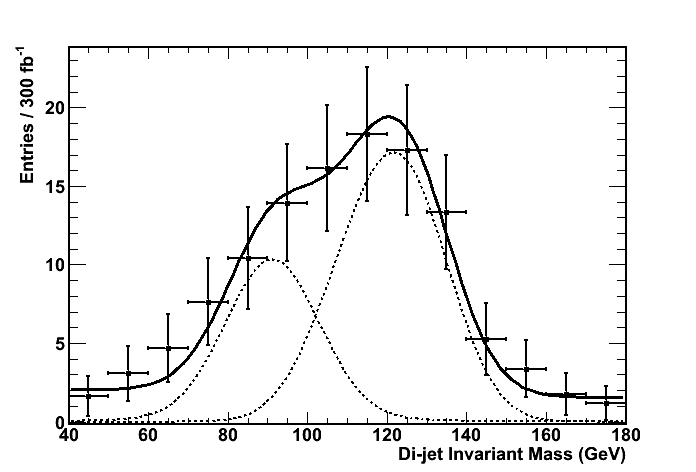} 
\caption{Di-jet invariant mass for 2 $b$ jets + lepton + missing energy $pp \rightarrow \chi^{\pm}_1 \chi^0_{2,3}$ events produced at 14~TeV. 
The result of the fit is shown by the continuous line, with the $Z$ and $h$ components represented by the dotted lines.}
\label{fig:MFitLHC}
\end{center}
\end{figure}

If a signal is observed in the $bb \ell$ + MET channel in the $13-14$~TeV LHC runs, some important measurements could be 
performed. In the case the $\tilde{\chi}^0_2$ and $\tilde{\chi}^0_3$ states are close in mass, as expected over a large part 
of the MSSM parameter space, they would be both produced at the LHC and may subsequently decay into $h \tilde{\chi}^0_1$ and 
$Z \tilde{\chi}^0_1$, as discussed above. In this case, the $b \bar b \ell$ + $E_{T}^{\mathrm{missing}}$ final 
state will receive contribution from both $h \to b \bar b$ and $Z \to b \bar b$, which can be resolved, at least on a 
statistical basis, from an analysis of the invariant mass of the reconstructed $bb$ system. The sensitivity of this analysis is 
assessed using an inclusive sample of $pp \rightarrow \tilde{\chi}^{\pm}_1 \tilde{\chi}^0_{2,3}$ generated at 14~TeV with an 
equivalent luminosity of 300~fb$^{-1}$ for a pMSSM point with the $\tilde{\chi}^{\pm}_1$, $\tilde{\chi}^{0}_2$ and $\tilde{\chi}^{0}_3$ 
at masses between 420 and 450~GeV. 
Events are reconstructed in the $b \bar b \ell$ + $E_T^{\mathrm{missing}}$ and those with one isolated lepton, two $b$-tagged 
jets and $E_T^{\mathrm{missing}} >$ 50~GeV are considered. The jet $b$-tagging efficiency is assumed to be 0.75. The invariant 
mass distribution of the $b$ jets is shown in Figure~\ref{fig:MFitLHC}. It receives contributions from the two dominant decays 
of the MSSM point chosen for simulation, $\chi^0_2 \rightarrow Z \tilde{\chi}^0_1$, $Z \rightarrow b \bar b$ and 
$\chi^0_3 \rightarrow h \tilde{\chi}^0_1$, $h \rightarrow b \bar b$ and the two sub-leading decays 
$\chi^0_3 \rightarrow Z \tilde{\chi}^0_1$, $Z \rightarrow b \bar b$ and $\chi^0_2 \rightarrow h \tilde{\chi}^0_1$,
 $h \rightarrow b \bar b$. From the branching fractions and cross sections of this specific pMSSM point, the fraction of $h$ in 
these decays is 0.73. 
The experimental di-jet mass resolution $\delta M/M$ for the $bb$ system at the $Z$ and $h$ masses is $\sim$0.13 from the {\tt Delphes} simulation, 
which agrees  with the performance obtained on full simulation for the LHC $H_{SM} \rightarrow b \bar b$ searches.
A $\chi^2$ fit is performed on the reconstructed invariant mass distribution to extract the fraction of decays into $h \tilde{\chi}$. The 
contribution from $Z$ and $h$ is modelled by Breit-Wigner distributions folded with a Gaussian resolution term and the combinatorial background 
is described by a first order polynomial. The $Z$ and $h$ fractions and the background parameters are left free in the fit. The result gives a 
fraction of decays with an $h$ boson of $0.65 \pm 0.09$, which agrees with the generated values within the statistical uncertainty. With the 
tenfold increase in data statistics expected by the HL-LHC program, the relative statistical uncertainty should be reduced from 14\% to $\sim$4\%, 
if the event reconstruction and $b$-tagging capabilities are maintained in the higher pile-up environment of the HL-LHC.

\subsection{$e^+e^-$ Linear Collider}

An $e^+e^-$ collider of sufficient energy and luminosity is very well suited for reconstructing two-body decays of neutralinos, pair produced 
in the process $e^+e^- \to \tilde{\chi}^0_i \tilde{\chi}^0_j$, including those decaying into $h$ and $Z$ bosons. In particular, since the energy 
of the produced $\tilde{\chi}^0_i$ is known, apart from beam radiation effects, the energy distribution of the reconstructed bosons can be used 
to precisely determine the $\tilde{\chi}^0_i$ mass. However, pair production cross sections are typically small.
The production of neutralino pairs in $e^+e^-$ collisions is mediated by $s$--channel $Z$ boson exchanges as well as $t$ and $u$--channel  
left-- and right--handed $\tilde e$  exchange. If the sleptons are assumed to be very heavy or the produced neutralinos are 
higgsino--like (and thus the coupling $\tilde{\chi}^0 \tilde e e$ is negligibly small)  only the $s$--channel $Z$ boson exchange is relevant.  
In fact, the neutralino cross sections are rather small, even in the case of $\tilde{\chi}_1^0 \tilde{\chi}_2^0$ production where the phase space 
is more favourable. The only exception is when both $\tilde{\chi}_{1}^0$ and $\tilde{\chi}_2^0$ are higgsino--like and have maximal couplings 
to the $Z$ boson.  Also the cross sections with identical neutralinos are extremely small. Due to Fermi statistics, the neutralinos are produced in 
$p$--waves and therefore suppressed at threshold and, in the case of  gauginos or higgsinos, the cross section is further suppressed by 
the couplings ${\cal Z}_{ii} \propto N_{i3}^2 - N_{i4}^2$. The cross section for mixed production of  
$\tilde{\chi}_1^0 \tilde{\chi}_3^0$, $\tilde{\chi}_2^0 \tilde{\chi}_3^0$ and $\tilde{\chi}_3^0 \tilde{\chi}_4^0$ 
are significant, except in the case where one of the neutralinos is a pure gaugino, which leads to suppressed 
$Z \tilde{\chi} \tilde{\chi}$ couplings. 

\begin{figure}[h!]
\begin{center}
\includegraphics[width=0.35\textwidth]{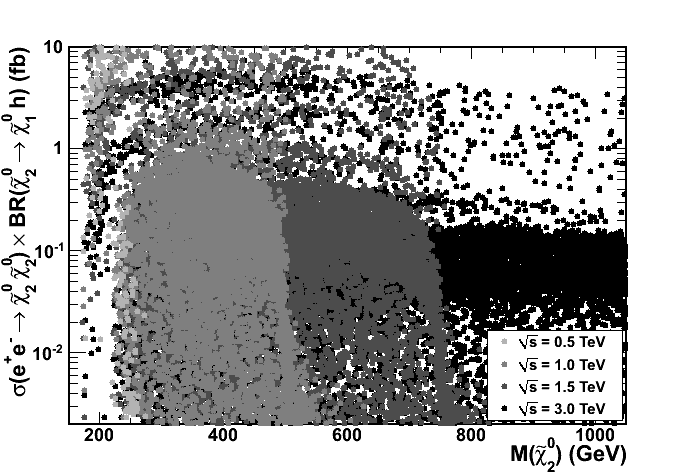} \\
\includegraphics[width=0.35\textwidth]{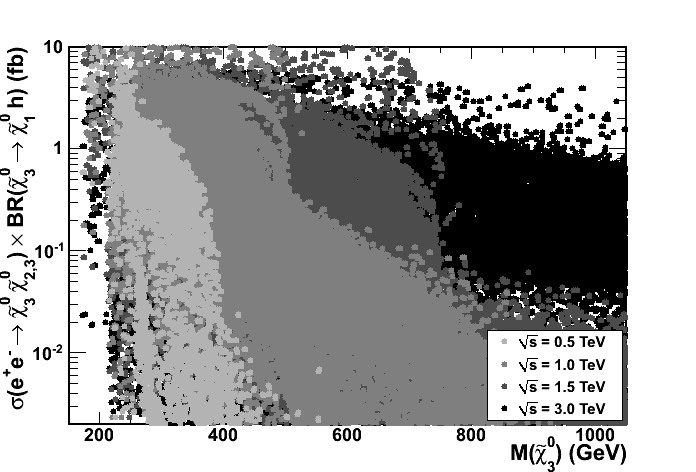} 
\caption{Product of neutralino pair production cross section and neutralino 
decay branching fraction into $h \tilde{\chi}^0_1$ as a function of the $\tilde{\chi}^0_2$ 
(upper) and $\tilde{\chi}^0_3$ (lower) masses at $\sqrt{s}$ = 0.5, 1, 1.5 and 3~TeV show by the dots in graded tones of grey.}
\label{fig:brxs}
\end{center}
\end{figure}

The product of the pair production cross sections and the $\tilde{\chi}^0_{2,3} \to h \tilde{\chi}^0_1$ branching fractions is 
shown in Figure~\ref{fig:brxs} as a function of the neutralino mass, for the four $\sqrt{s}$ energy values (0.5, 1, 1.5 and 3~TeV) 
chosen for this study.

\begin{figure}[h!]
\begin{center}
\includegraphics[width=0.35\textwidth]{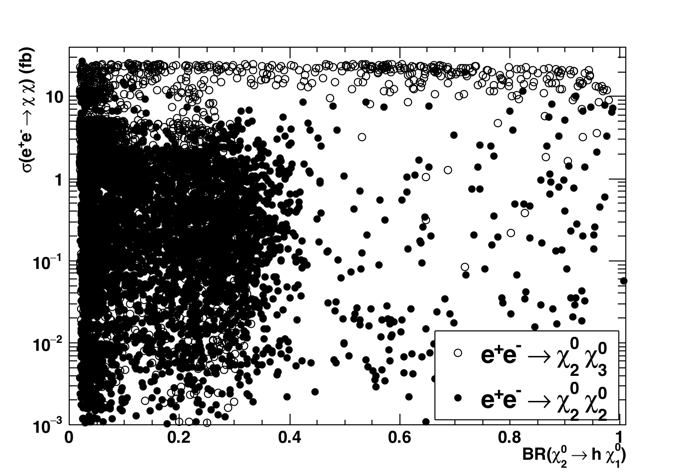}
\caption{Neutralino pair production cross section for 
$e^+e^- \to \tilde{\chi}^0_2 \tilde{\chi}^0_2$ (filled circles) and $\tilde{\chi}^0_2 \tilde{\chi}^0_3$ (open circles) 
at $\sqrt{s}$ = 1.5~TeV shown as a function of the $\tilde{\chi}^0_2 \to h \tilde{\chi}^0_1$ branching fraction for 
pMSSM points where the sum of the neutralino masses, $2 M_{\tilde{\chi}^0_2}$ and $M_{\tilde{\chi}^0_2} + M_{\tilde{\chi}^0_3}$ 
respectively, does not exceed 1.4~TeV.}
\label{fig:brxsn23}
\end{center}
\end{figure}

It is important to observe that, in the region of parameters where the 
$\tilde{\chi}^0_2 \to h \tilde{\chi}^0_1$ process is the dominant neutralino decay, the $\tilde{\chi}^0_2$ tends to decouple 
from the $Z$. This suppresses the $e^+e^- \to \tilde{\chi}^0_2 \tilde{\chi}^0_2$ pair production, leaving instead the 
$e^+e^- \to \tilde{\chi}^0_2 \tilde{\chi}^0_3$ process as the only significant neutralino production mechanism, 
when kinematically allowed. This correlation between pair production cross section and decay branching fractions is illustrated 
in Figure~\ref{fig:brxsn23} for a sample of accepted pMSSM points, where the sum of the masses of the pair-produced neutralinos 
is more than 100~GeV below the collision energy, to avoid threshold effects. It is evident that the region of large branching
fractions into $h$ bosons is dominated by larger $\tilde{\chi}^0_2 \tilde{\chi}^0_3$ production.

Chargino and neutralino pair production can be typically observed at a linear collider operating at a centre of mass energy of $\sqrt{s}$ 
for $\mu$ and $M_2$ values up to the kinematic limit of $\sqrt{s}/2$. This makes the sensitivity of a LC, already at 0.5~TeV and more
definitely at 1.0~TeV and above, highly complementary to that of the LHC at 14~TeV. Due to the relatively large mass splitting from the LSP
required for sensitivity in the $\tilde{\chi}^{\pm}_1 \tilde{\chi}^0_2$ production process, the parameter region where the LHC may detect
chargino - neutralino production starts at $M_2 >$ 250~GeV and $\mu >$ 250~GeV. The LC operating already at 0.5~TeV will completely cover
the complementary region of $M_2 <$ 250~GeV or $\mu <$ 250~GeV. The decay $\tilde{\chi}^0_2 \to Z \tilde{\chi}^0_1$ was studied as a benchmark
reaction at the ILC at $\sqrt{s}$ = 0.5~TeV using full simulation and reconstruction as part of the ILC LoIs~\cite{ild,sid}. In these analyses,
the $\tilde{\chi}^0_2$ mass of 217~GeV was determined with a statistical 
relative accuracy of 1\%. The decays $\tilde{\chi}^0_2 \to h \tilde{\chi}^0_1$ and $\tilde{\chi}^0_2 \to Z \tilde{\chi}^0_1$ have been
studied for CLIC at $\sqrt{s}$ = 3~TeV~\cite{Alster:2011he}. 
The $\tilde{\chi}^0_2 \to h \tilde{\chi}^0_1$ decay, followed by $h \to b \bar b$, leads to the signature $b \bar b b \bar b +$ missing 
energy final state. The $\tilde{\chi}^0_2$ decay can be isolated in an inclusive SUSY sample, where its mass and the $h$ yield are 
determined with a relative statistical accuracy of 4\% and 5\%, respectively.

\begin{figure}[h!]
\begin{center}
\begin{tabular}{cc}
\includegraphics[width=0.23\textwidth]{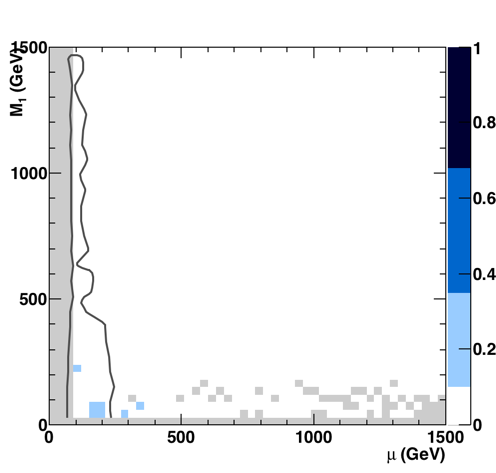} &
\includegraphics[width=0.23\textwidth]{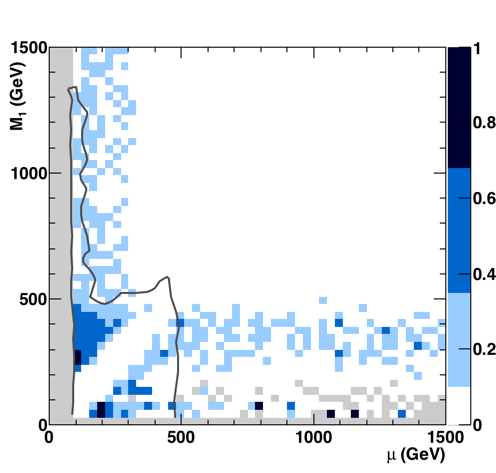} \\
\includegraphics[width=0.23\textwidth]{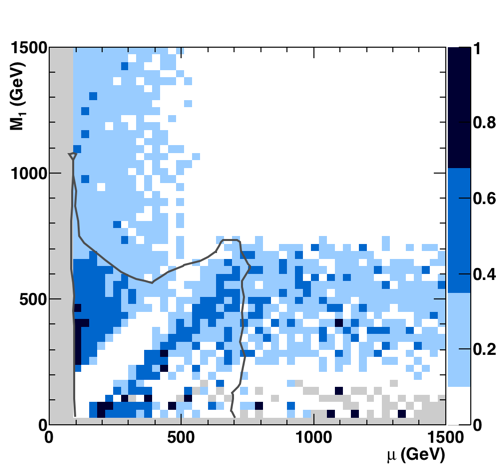} &
\includegraphics[width=0.23\textwidth]{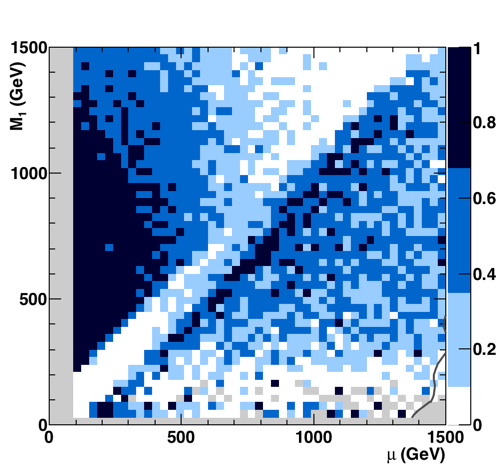} \\
\end{tabular}
\caption{Fraction of accepted pMSSM points in the $[\mu - M_1]$ plane giving at least 25 events in the chargino and neutralino pair production
channels with $\tilde{\chi}^0_{2,3} \to h \tilde{\chi}^0_1$ for 0.5~ab$^{-1}$ of $e^+e^-$ data at $\sqrt{s}$ = 0.5~TeV (upper left),  1~ab$^{-1}$ at $\sqrt{s}$ = 1~TeV (upper right), 
2.0~ab$^{-1}$ at $\sqrt{s}$ = 1.5~TeV (lower left) and 3~ab$^{-1}$ at $\sqrt{s}$ = 3~TeV (lower right). The grey lines indicate the regions
where more than 68\% of the accepted pMSSM points give at least 25 signal events in all chargino and neutralino pair production
channels.}
\label{fig:M1MuLC}
\end{center}
\end{figure}

With an expected integrated luminosity of $0.5 - 3$~ab$^{-1}$ a high energy linear collider can observe these decays over a 
significant part of the kinematically allowed region of the parameter space. This is quantified here by the fraction of the 
pMSSM points yielding at least 25 signal events for an integrated luminosity of 0.5~ab$^{-1}$ for $\sqrt{s} =$ 0.5~TeV and 
of 1~ab$^{-1}$ for 1~TeV,  2~ab$^{-1}$ for 1.5~TeV and 3~ab$^{-1}$ for 3~TeV, over the $[\mu, M_1]$  parameter space, as shown 
in Figure~\ref{fig:M1MuLC}, where the sensitivity to the decay into $h$ is compared to that of inclusive neutralino and chargino
decays. 

The background for these decays with two $h$ or a $h$ and a $Z$ bosons and large missing energy are expected to be small and
this justifies our simple criteria of event counting to obtain the regions of sensitivity in the $[\mu - M_1]$ plane.
These backgrounds have been studied in more detail at 3~TeV, as discussed below.

Once a signal is observed, the yield of $h$ bosons can be determined in an inclusive SUSY sample. We study it for the specific case of a MSSM 
model having $\tilde{\chi}^0_1$ with mass of 340~GeV, $\tilde{\chi}^0_2$ with mass of 643~GeV decaying predominantly into $h \tilde{\chi}^0_1$ 
and $\tilde{\chi}^{\pm}_1$ with mass of 643~GeV decaying exclusively into $W^{\pm} \tilde{\chi}^0_1$ and $M_h$ = 126.1~GeV, in 
3~TeV $e^+e^-$ collisions. An inclusive sample of SUSY events, corresponding to 0.5~fb$^{-1}$ of integrated luminosity, is generated using 
{\tt Pythia} and events are fully simulated using {\tt Geant-4}~\cite{g4} and reconstructed following the same analysis procedure 
discussed in~\cite{Alster:2011he}. Events are pre-selected requiring a visible energy  $0.08 \sqrt{s} < E_{tot} < 0.6 \sqrt{s}$, an energy 
in charged particles larger than 150~GeV, transverse energy larger than 200~GeV and jet multiplicity 2$\le N_{jets} <$5. Jet clustering is 
performed using the Durham jet algorithm~\cite{Catani:1991hj}, with $y_{cut}$ = 0.0025, on the reconstructed particle flow objects of the 
Pandora particle flow package~\cite{Thomson:2009rp}. Selected events are clustered into four jets and the di-jet invariant mass for all the 
three possible pairings is computed. The jet pairing minimising the difference between the di-jet invariant masses is selected, provided the 
mass difference is below 20~GeV. The fraction of background non-resonant events is obtained from the di-jet mass side-bands 20 $< E_{jj} <$ 60~GeV 
and 140 $< E_{jj} <$ 200~GeV. The fraction of $W^+W^-$, $ZZ$ and $hh$ events in the inclusive 4-jet SUSY events 
is extracted by a $\chi^2$ fit to this di-jet mass distribution. The $W^{\pm}$ and $Z$ mass peaks are parametrised as Breit-Wigner functions 
convoluted with a Gaussian term describing the experimental resolution. The mass and width values of the Breit-Wigner functions are fixed to 
their generated values, while the total area and the width of the Gaussian resolution terms are left free in the fit. 
\begin{figure}[h!]
\begin{center}
\includegraphics[width=0.35\textwidth]{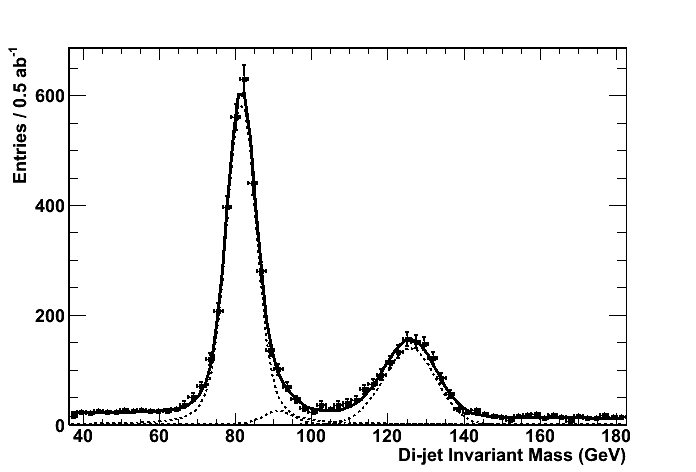} \\
\includegraphics[width=0.35\textwidth]{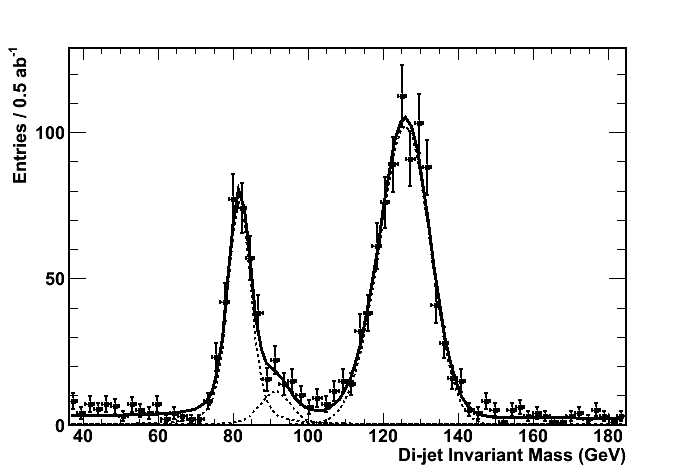} 
\caption{Di-jet invariant mass in inclusive 4-jet + missing energy SUSY events produced in $\sqrt{s}$=3~TeV $e^+e^-$ collisions for 
all selected events (upper) and $b$-tagged events (lower). The result of the fit is shown by the continuous line with the individual 
$W$, $Z$ and $h$ components represented by the dotted lines.}
\label{fig:MFit}
\end{center}
\end{figure}
The $h$ peak is modelled as the sum of two Gaussian curves, one representing the correctly reconstructed signal events, centred at the nominal $M_h$ value, the second describing decays where the mass has a lower reconstructed value due to semi-leptonic $b$ decays. The central value, width and fraction of events in this second Gaussian is extracted by a fit to a pure sample 
of decays into bosons and fixed in the fit, while the Gaussian width of the main peak is kept free.
The yield of $h$ bosons, with mass 126.1~GeV, is extracted by a $\chi^2$ fit to the di-jet mass distribution of events in 4-jet final 
states, shown in Figure~\ref{fig:MFit} for all di-jets and for those passing di-jet b-tagging based on the ZVTOP algorithm~\cite{zvtop} 
implemented in the {\tt LCFIVertex} package~\cite{Bailey:2009ui}. We measure the fraction of $h$ bosons to be 0.269$\pm$0.013 and that of $Z$ 
bosons to be 0.037$\pm$0.016, after the non-resonant background subtraction, which compares well to the original values of 0.290 and 0.025, 
respectively, of the generated events. This specific example shows the feasibility to accurately determine the yield of $h$ bosons in SUSY 
particle decays in the data of an $e^+e^-$ collider operating at sufficient energy.

\section{Conclusions}

The observation of a Higgs particle with a mass of $\simeq$126~GeV promotes the exploration of its possible role in the decay of
weakly-interacting particles  in the context of supersymmetry. Neutralino decays offer a prime opportunity, since the
$\tilde{\chi}^0_{2,3} \to h \tilde{\chi}^0_1$ process provides the largest yield of $h$ bosons in SUSY decays in the MSSM and this
decay channel is highly complementary to channels leading to multi-lepton final states, such as $Z$ and $\tilde \ell \ell$.
The typical values of cross sections times decay branching fractions to $bb \ell$ final state are of the order $0.01 - 1$~fb in 8~TeV
$pp$ collisions and a factor of about three larger at 14~TeV for $M_{\chi^0_{2,3}} \le$ 600~GeV. The LHC experiments have just started
probing these processes. Extrapolating to 13-14~TeV the preliminary results obtained in the first searches with the $bb \ell$+MET channel
at 8~TeV shows that the increase in cross section and available statistics should make possible to probe a significant region of the
MSSM parameter space. At 13-14~TeV, the 
neutralino decays into Higgs should account for more than half of the pMSSM points for which the electroweak $\tilde{\chi}^{\pm}_1 \tilde{\chi}^0_{2,3}$ production channels have sensitivity. 
The points accessible through the $h$ channel belong to regions complementary to those covered by searches with multi-leptons. If a signal is observed, 
then the fraction of $h$ bosons produced in neutralino decays could be measured to a statistical accuracy of better than 10\% at the HL-LHC. 
An $e^+e^-$ linear collider of sufficient energy, $\sqrt{s} \ge$ 1~TeV, and luminosity, yielding $\ge$1.0-3.0~ab$^{-1}$ of data, can study 
chargino neutralino pair production in regions of the parameter space highly complementary to the LHC at 14~TeV. Further, it can study $h$ 
production in neutralino decays for masses up to the kinematical limit for pair production, obtaining $\cal{O}$(1\%) precision on the mass 
and $\cal{O}$(5\%) on the yield of Higgs bosons from neutralino decays in inclusive multi-jet SUSY events.

\section*{Acknowledgements}

We thank A.~Djouadi for extensive discussion and A.~Canepa and Z.~Gecse for useful suggestions.
M.B. wishes to thank the Galileo Galilei Institute for Theoretical Physics for the hospitality
and INFN for partial support during the development of this study.

\end{document}